\documentclass[aps,prl,longbibliography,twocolumn,superscriptaddress,showpacs]{revtex4-1}

\usepackage{CJK}

\usepackage{graphicx}

\usepackage{amsmath}
\usepackage{dcolumn}

\usepackage{bm}
\usepackage[normalem]{ulem}
\usepackage{color}
\usepackage{hyperref}

\usepackage{epstopdf}
\newcommand{\bew}{\begin{widetext}}
\newcommand{\ew}{\end{widetext}}

\newcommand{\ii}{{\rm i}}

\newcommand{\bq}{\mathbf{q}}

\newcommand{\br}{\mathbf{r}}

\newcommand{\bbr}{\mathbf{r}}

\newcommand{\beq}{\begin{equation}}
\newcommand{\eeq}{\end{equation}}
\newcommand{\beqn}{\begin{eqnarray}}
\newcommand{\eeqn}{\end{eqnarray}}

\newcommand{\dd}{{\rm d}}
\newcommand{\ee}{{\rm e}}

\begin{document}
\title{Roughening of two-dimensional interfaces  in nonequilibrium phase-separated systems}
\author{John Toner}
\email{jjt@uoregon.edu}
\affiliation{Department of Physics and Institute of Theoretical
Science, University of Oregon, Eugene, OR $97403^1$}
\date{\today}

	\begin{abstract}
	{I show that non-equilibrium two-dimensional interfaces between  three dimensional phase separated fluids exhibit a peculiar "sub-logarithmic" roughness. Specifically, an interface of lateral extent $L$ will fluctuate vertically (i.e., normal to the mean surface orientation) a typical RMS distance $w\equiv\sqrt{\langle |h(\br,t)|^2\rangle} \propto [\ln{(L/a)}]^{1/3}$ (where $a$ is a microscopic length, and  $ h(\br,t)$ is the height of the interface at two dimensional position $\br$ at time $t$). In contrast, the roughness of equilibrium two-dimensional interfaces between  three dimensional  fluids, obeys $w \propto [\ln{(L/a)}]^{1/2}$. The exponent $1/3$ for the active case is exact. In addition, the characteristic time scales $\tau(L)$ in the active case scale according to $\tau(L)\propto L^3 [\ln{(L/a)}]^{1/3}$, in contrast to the simple $\tau(L)\propto L^3$ scaling found in equilibrium systems with conserved densities and no fluid flow.}
	\end{abstract}
	
\maketitle

The KPZ equation\cite{KPZ} is an excellent illustration of the novelty of non-equilibrium statistical mechanics. It describes moving interfaces (e.g., the surface of a crystal growing into a supersaturated vapor), and its solution reveals that the behavior of such non-equilibrium interfaces is very different from that of their equilibrium counterparts\cite{EW}.

Recently, it was recognized\cite{besse} that non-equilibrium interfaces in phase separated systems with a  {\it conserved} density field belong to an entirely new universality class, which \cite{besse} dubbed "the $|\bq|KPZ$ equation". They studied this universality class using the dynamical renormalization group (DRG), and estimated its scaling exponents in one dimension (1d) (that is, for a one-dimensional interface between a pair of two dimensional media) using a $2-\epsilon$ expansion. Their results showed conclusively that the 1d problem was indeed a different universality class from the 1d KPZ equation.

In this paper, I show that this also holds in $d=2$ (2d). More specifically, the $|\bq|KPZ$ equation also belongs to a new universality class, exhibiting scaling behavior different from {\it both} the 2d KPZ equation, and the equilibrium Edwards-Wilkinson model\cite{EW}.

To be more quantitative, I find that relaxation times $\tau(L)$of interface fluctuations on a length scale $L$ in the $|\bq|KPZ$ equation  scale according to 
\beq
\tau(L)\propto L^3 [\ln{(L/a)}]^{1/3} \,,
\label{tscale}
\eeq
 in contrast to the simple $L^3$ scaling found in equilibrium systems with a conserved density field. 
This scaling is  also very different from that found in the 2d KPZ equation\cite{KPZ}.

In addition, the characteristic width of an interface of lateral extent $L$ for the $|\bq|KPZ$ equation grows more slowly with $L$ than in both the equilibrium case\cite{EW}, and the 2d KPZ equation\cite{KPZ}. I find for the $|\bq|KPZ$ equation:
\beq
 w\equiv\sqrt{\langle |h(\br,t)|^2\rangle} \propto [\ln{(L/a)}]^{1/3} \,,
\label{width}
\eeq
where $a$ is a microscopic length, and  $h(\br,t)$ is the height of the interface at two dimensional position $\br$ at time $t$. 

Both the exponents $1/3$ in the logarithms in (\ref{tscale}) and (\ref{width}) are {\it exact}, as is the power $3$ on the $L^3$ in (\ref{tscale}).

I'll now derive these results, starting with the model of \cite{besse}, and the examining first its linear, and then its non-linear behavior.

  {\it Model.---}
 In Fourier space, the equation of motion (EOM) for an active growing interface with between two regions of different densities with a conserved density field was derived in \cite{fausti, besse}, and reads
 \beqn
-i\omega h(\bq,\omega)&=&-\sigma q^3 h(\bq,t)+{\lambda\over2}q \mathcal{F}_{\bq,\omega}\left[
|\nabla h(\br,t)|^2\right]+f(\bq,t)  \,.
\nonumber\\
\label{eq:u_eom}
\eeqn
Here $q\equiv|\bq|$,  the symbol $\mathcal{F}_{\bq, \omega}$ represents the ${\bq, \omega}$'th Fourier component, i.e.,
\beqn
\mathcal{F}_{\bq, \omega}\left[h(\br,t)\right]\equiv{1\over (2\pi)^{d/2}}
\int\dd^dr\, \ee^{\ii(\bq \cdot \bbr-\omega t)}h(\br,t)\,,\nonumber\\
\eeqn
and the Gaussian "blue noise" $f(\bq,t)$ has zero mean and variance given by
\beq
\langle f(\bq,t)f(\bq',t')\rangle=
2Dq\delta^d(\bq+\bq')\delta(t-t')\,.
\label{blue}
\eeq

The Brillouin zone of all Fourier transforms has an ultraviolet cutoff $\Lambda$; that is, wavevectors $\bq$ are restricted to the interior of a hypersphere: $|\bq|<\Lambda$.



\noindent{\it Linear theory}--- The linearized version of the EOM \eqref{eq:u_eom}, which can be obtained simply by dropping the non-linear $\lambda$ term,was first studied by Bray et al\cite{Bray}.  It can be solved for $h(\bq,\omega)$ in terms of the random force $f(\bq,\omega)$:
\beq
 h(\bq, \omega)=G(\bq, \omega) f(\bq, \omega) \,,
\label{hsol1.0}
\eeq
where  the ``propagator" 
\beq
G(\bq, \omega) \equiv {1\over -i\omega+\sigma q^3} \,.
\label{Gdef}
\eeq

The existence of a pole in this propagator at
\beq
\omega=-i\sigma q^3
\label{decay rate}
\eeq
shows that excitations of wavevector $\bq$ decay at a rate $-i\sigma q^3$, according to the linear theory. We will see later that this scaling changes when the $\lambda$ non-linearity is taken into account.

Autocorrelating this with itself, and using the expression (\ref{blue}) for the random force correlations gives
\bew
\beqn
\langle h(\bq, \omega) h(\bq', \omega')\rangle&=&G(\bq, \omega)G(\bq', \omega')\langle f(\bq, \omega) f(\bq', \omega')\rangle=2DqG(\bq, \omega)G(\bq', \omega')\delta^d(\bq+\bq')\delta^d(\omega+\omega') 
\nonumber\\
&=&2DqG(\bq, \omega)G(-\bq, -\omega)\delta^d(\bq+\bq')\delta^d(\omega+\omega') \equiv C(\bq, \omega)\delta^d(\bq+\bq')\delta^d(\omega+\omega') \,,
\label{corr1}
\eeqn
\ew
with
\beq
C(\bq,\omega)= {2D q
\over(\omega^2+\sigma^2 q^6)} \,.
\label{CKPZ}
\eeq
Integrating this expression over the frequencies $\omega$ and $\omega'$ gives the {\it spatially} Fourier  transformed {\it equal time} correlation function:
\beqn
\langle h(\bq, t) h(\bq',t)\rangle&=&C_{\rm ET}(\bq)\delta^d(\bq+\bq') 
\label{corret}
\eeqn
with
\beq
C_{\rm ET}(\bq)=\int{d\omega\over2\pi} {2D q
\over(\omega^2+\sigma^2 q^6)}={D\over\sigma q^2} \,.
\label{et final}
\eeq

This correlation function diverges like $1/q^2$ as $q\to0$. This means that the surface will, according to the linearized theory, be algebraically ``rough" in spatial dimensions $d<2$, logarithmically rough in $d=2$, and ``smooth" in $d>2$. Specifically,
\bew
\beq
\langle (h(\br, t))^2\rangle=\int{d^dq\over(2\pi)^d}\int{d^dq\over(2\pi)^d}\langle h(\bq, t) h(\bq',t)\rangle=\int{d^dq\over(2\pi)^d}C_{\rm ET}(\bq)={D\over\sigma (2\pi)^d}\int{d^dq\over q^2}
\propto\left\{
\begin{array}{ll}
L^{2-d},&d<2\,,\\
\ln\left({ \Lambda L\over2\pi}\right),&d=2\,,\\
{\rm finite}\,{\rm (independent}\, {\rm of}\,L),&d>2 \,,
\label{hscale1}
\end{array}\,
\right.
\eeq
\ew
where the system size $L$ acts as an infrared cutoff to the integral over $q$.
In \cite{besse}, it was shown that the scaling of both the fluctuations (\ref{hscale1}) and the relaxation times (\ref{decay rate}) change for one dimensional interfaces ($d=1$) due to the effect of the non-linear $\lambda$ term in (\ref{eq:u_eom}). In the next section, I will show that this is also true for two dimensional interfaces.

{\it Nonlinear regime \& DRG analysis.---}
 I now turn to the full, non-linear EOM of $h$ (\ref{eq:u_eom}). This was studied   by \cite{besse}using the dynamical renormalization group (DRG). They found the parameters in the renormalization group obeyed the recursion relations
 \begin{subequations}
\label{rr}
\begin{align}
{\dd\ln\sigma\over\dd\ell}&=z-3+g \,,\label{sigma}\\
{\dd\ln{\lambda}\over\dd\ell}&=z-3+\chi\,,\label{lambda}\\
{\dd \ln D\over\dd\ell}&=z-2\chi-d-1\,,\label{fl_D}
\end{align}
\end{subequations}
where I've defined the dimensionless coupling
\beq
g\equiv {S_dD\lambda^2\over8(2\pi)^d\sigma^3}\,,
\label{g_def}
\eeq
with $S_d$ being the surface area of a $d$-dimensional unit sphere.\cite{diff g}

In (\ref{sigma})-(\ref{fl_D}), $z$ is the "dynamical" exponent for the rescaling of time under the RG, and  $\chi$ is the "roughness" exponent for the rescaling of the field $h(\br,t)$ in real space. I will later  make specific convenient choices for these arbitrary exponents.

From \eqref{rr}, one can construct a closed recursion relation
for $\ln g$:
\beq
{\dd \ln g\over\dd\ell}=\epsilon-3g\,,
\label{Flow_lng}
\eeq
where $\epsilon\equiv2-d$. This can obviously be turned into a closed recursion relation for $g$ itself:
\beq
{\dd g\over\dd\ell}=\epsilon g-3g^2\,.
\label{gen Flow g}
\eeq
This recursion relation and its physical consequences were analyzed for $d<2$ by \cite{besse}. Here, I wish to focus on $d=2$, where $\epsilon=0$, and the recursion relation (\ref{gen Flow g}) becomes
\beq
{\dd g\over\dd\ell}=-3g^2\,.
\label{gen Flow g}
\eeq

The solution of this recursion relation is straightforward:
\beq
g(\ell)={g_0\over1+3g_0\ell}\approx{1\over3\ell} \,,
\label{gsol}
\eeq
where $g_0\equiv g(\ell=0)$ is the "bare" value of the dimensionless coupling $g$, and the last, approximate equality holds at large RG time $\ell$.

This solution is, strictly speaking, only valid if the bare dimensionless coupling $g_0$ is small, since only in that case will the recursion relation (\ref{gen Flow g}), which was derived assuming that $g(\ell)\ll1$, hold.

However, one can make a stronger statement: as long as the bare $g(\ell)$ lies in the basin of attraction of the RG fixed point at $g=0$, $g(\ell)$ will eventually flow to sufficiently small values that  (\ref{gen Flow g}) {\it does} hold. Once that happens, $g(\ell)$ is guaranteed to flow to even smaller values, and, eventually, to exhibit the asymptotic behavior $g(\ell)\approx1/(3\ell)$ displayed in equation (\ref{gsol}). Thus, for any such initial condition (which may, for all we know, include {\it all} initial conditions), the aforementioned asymptotic behavior will hold.

One might be mislead into thinking that, since the dimensionless coupling $g$ that determines the importance of the nonlinearity $\lambda$ vanishes upon renormalization, the effect of that non-linearity vanishes at long length and time scales. However, because the vanishing of $g(\ell)$ is so slow (i.e., proportional to $1/\ell$, rather than exponential), this proves not to be the case. In fact, as I'll show in a moment, the vanishing of $g$ in two dimensions is {\it so} slow that it leads to {\it infinite} renormalization of the effective "surface tension" $\sigma$, which in turns leads to "anomalous hydrodynamics": that is, different scaling laws for the long-distance, long-time dynamics of the system than those predicted by a purely linear theory which ignored the nonlinearity $\lambda$.

This situation is very analogous to that in equilibrium three dimensional smectic liquid crystals\cite{pelc}, which also have a dimensionless coupling constant ( in that case associated with the anharmonic terms in the smectic elastic Hamiltonian) which vanishes like $1/\ell$ as RG "time" $\ell\to\infty$. In that case, this leads to "anomalous elasticity", in which both the elastic response of the smectic, and its fluctuations, scale differently at large length scales than predicted by a purely harmonic elastic theory.

To see this in our problem, I'll begin by using the solution (\ref{gsol}) for $g(\ell)$ in the recursion relation (\ref{sigma}) for the surface tension $\sigma$, and solving the resultant equation for the renormalized $\sigma(\ell)$. At large $\ell$, where the asymptotic form given in equation (\ref{gsol}) holds, the recursion relation for $\sigma$ becomes
\beq
{\dd\ln\sigma\over\dd\ell}\approx z-3+{1\over3\ell} \,.
\label{sigma2}
\eeq
Since the time rescaling exponent $z$ is arbitrary, I will make the simplifying choice $z=3$. With this choice, the solution of (\ref{sigma2}) is
\beq
\sigma(\ell)=\sigma_0\ell^{1/3} \,,
\label{sigmasol}
\eeq
where $\sigma_0$ is a non-universal constant set by the bare values of the parameters.
Note that $\sigma(\ell)$ diverges as $\ell\to\infty$; this is the origin of the anomalous hydrodynamics in two dimensions. 

To see this,
I will now use the trajectory integral matching method\cite{Traject} to compute the correlations of the fluctuating field $h$. This approach implies that the spatio-temporally Fourier transformed correlation function  $C(\bq, \omega; \sigma_0, D_0, \lambda_0)\equiv\langle \left|h(\vec{q}, \omega)\right|^2\rangle$ in our original system ("original" in the sense of "before renormalization") is related to that of the renormalized system through the relation:
\beqn
C(\bq, \omega; \sigma_0, D_0, \lambda_0)&=&e^{[2\chi+z+d]\ell}C(e^\ell\bq, e^{z\ell}\omega; \sigma(\ell), D(\ell), \lambda(\ell))\,.
\nonumber\\
\label{cqwti}
\eeqn
This relation (\ref{cqwti})  holds for any value of the renormalization group "time"
$\ell$. I am completely free to choose that "RG time" to be anything I want. One convenient choice is
\beq
e^\ell q=\Lambda \Rightarrow \ell=\ln\left({\Lambda\over q}\right) \,.
\label{ellchoice}
\eeq
With this choice,  the magnitude of the wavevector $e^\ell \bq$ on the right hand side of (\ref{cqwti}) is $\Lambda$, the ultraviolet cutoff. At such a large value of wavevector, the correlation function $C(\Lambda, e^{z\ell}\omega; \sigma(\ell), D(\ell), \lambda(\ell))$ should be given accurately by the linear theory, with the renormalized values $\sigma(\ell)$ and $D(\ell)$ of $\sigma$ and $D$. This is because  the perturbative corrections to the linear results only become important at {\it small} $\bq$. Furthermore,  $g(\ell)$, which determines the importance of the non-linearity, is very small at large $\ell$. Note that the choice (\ref{ellchoice}) {\it will} be large  at small $\bq$ - indeed, it diverges as $\bq\to{\bf 0}$ - so this approximation becomes asymptotically exact as $\bq\to{\bf 0}$.

Therefore, I have
\beq
C(e^\ell\bq, e^{z\ell}\omega; \sigma(\ell), D(\ell), \lambda(\ell))={2D(\ell)e^\ell q\over e^{2z\ell}\omega^2+\sigma^2(\ell)\Lambda^6} \,,
\label{cren}
\eeq
where I've used the fact that $e^\ell|\bq|=\Lambda$.

I have already obtained $\sigma(\ell)$  in equation (\ref{sigmasol}) by making the convenient choice of dynamical exponent $z=3$. In addition, I want to focus on spatial dimension $d=2$. Finally,  I'm also free to choose the field rescaling exponent;  the  convenient choice is $\chi=0$. With these values for $d=2$, $\chi=0$, and $z=3$, the recursion relation (\ref{fl_D})  for the noise strength $D(\ell)$ becomes simply ${\dd \ln D\over\dd\ell}=0$.  This trivially implies that $D(\ell)=D_0$, its bare value. Using this value for $D$,  the expression (\ref{ellchoice}) for the RG time $\ell$, and the result (\ref{sigmasol}) for the renormalized "surface tension" $\sigma(\ell)$ in the expression (\ref{cren}) for the correlation function in the renormalized system, and using the result in the trajectory integral matching formula  (\ref{cqwti}), I obtain my final expression for the spatiotemporally Fourier transformed correlation function of the original (i.e., the experimental) system:
\beq
C(\bq,\omega)= {2D_0 q
\over\bigg(\omega^2+\sigma_0^2 \left[\ln\left({\Lambda\over q}\right)\right]^{2/3}q^6\bigg)} \,.
\label{CKPZNL}
\eeq

The  
alert reader will recognize this expression as being identical to the linear result (\ref{CKPZ}), {\it except} for the replacement of the bare 
surface tension $\sigma$ of the linear theory with a {\it wavevector-dependent}, "renormalized" $\sigma_R(\bq)$ given by
\beq
\sigma_R(\bq)\equiv\sigma_0 \left[\ln\left({\Lambda\over q}\right)\right]^{1/3} \,,
\label{sigmarenq}
\eeq
which {\it diverges} as $\bq\to0$.

The existence of a pole in the correlation function (\ref{CKPZNL}) at
\beq
\omega=-i\sigma_R(\bq) q^3=-i\sigma_0 \left[\ln\left({\Lambda\over q}\right)\right]^{1/3} q^3
\label{decay rate NL}
\eeq
shows that excitations of wavevector $\bq$ decay at a rate $\sigma_R(\bq)q^3=\sigma_0 \left[\ln\left({\Lambda\over q}\right)\right]^{1/3} q^3$ in the full, non-linear theory. This is different scaling from that predicted in equation (\ref{decay rate}) by the linear theory; this is the change in the scaling of the relaxation rate referred to earlier.

This change in relaxation rates  leads to a different scaling of the real space fluctuations as well.
With  the result (\ref{CKPZNL}) for the spatiotemporally Fourier transformed correlation function of the full nonlinear theory in hand, we can proceed as we did in the linear theory.
Integrating this expression over the frequencies $\omega$ and $\omega'$ gives the {\it spatially} Fourier  transformed {\it equal time} correlation function:
\beqn
\langle h(\bq, t) h(\bq',t)\rangle&=&C_{\rm ET}(\bq)\delta^d(\bq+\bq') 
\label{corret}
\eeqn
with
\beq
C_{\rm ET}(\bq)=\int{d\omega\over2\pi} {2D q
\over(\omega^2+\sigma_R(\bq)^2 q^6)}={D\over\sigma_R(\bq) q^2} \,,
\label{et final}
\eeq
and $\sigma_R(\bq)$ the renormalized surface tension given by (\ref{sigmarenq}), which I remind the reader diverges as $\bq\to{\bf 0}$.

This correlation function  diverges like $1/\bigg(q^2\left[\ln\left({\Lambda\over q}\right)\right]^{1/3}\bigg)$ as $q\to0$. This means that the roughness of the surface will now be given by
\bew
\beq
\langle (h(\br, t))^2\rangle=\int{d^2q\over(2\pi)^2}{D\over\sigma_R(\bq)  q^2}={D_0\over2\pi\sigma_0}\int_{L^{-1}}^\Lambda {dq\over q \left[\ln\left({\Lambda\over q}\right)\right]^{1/3}}={3D_0\left[\ln\left({\Lambda L}\right)\right]^{2/3}\over4\pi\sigma_0} \,.
\label{roughfin}
\eeq
\ew
The rms roughness $w$ of the surface is the sqaure root of this expression:
\beq
w=\sqrt{3D_0\over4\pi\sigma_0}\left[\ln\left({\Lambda L}\right)\right]^{1/3} \,.
\label{width2}
\eeq
So the surface is much less rough, in the limit of large system size $L$,  than the $\sqrt{\ln\left({\Lambda L}\right)}$ roughness predicted by the linear theory.

These results were all derived using the recursion relations (\ref{rr}) of \cite{besse}. Those recursion relations are only valid to leading order in the dimensionless coupling $g$. However, since, as shown by the solution (\ref{gsol}), $g(\ell\to\infty)\to0$, my results actually become exact in the long-distance, long time limit.

In conclusion, I have shown that non-equilibrium two dimensional interfaces in phase separated systems with a  {\it conserved} density field belong to a new universality class, different from both equilibrium systems (whether with or without conserved densities) and the nonequilibrium KPZ equation. I've also shown that such interfaces have logarithmically faster relaxation than in equilibrium, and exhibit widths that diverge with system size as a smaller power of the logarithm of the system size than in  equilibrium.

\begin{acknowledgments}
I thank Marc Besse for introducing me to  reference \cite{besse}.
\end{acknowledgments}


\begin{thebibliography}{16}
\bibitem{diff g} Note that this definition of $g$ differs from that of reference \cite{besse} by a constant numerical factor of ${S_d\over 8(2\pi)^d}$, where $S_d$ is the surface hyper-area of a $d$-dimensional hypersphere. For the special case of $d=2$, the "hypersphere" is just a  circle, and $S_d=S_2=2\pi$ is just the circumference of a unit circle.
\end{thebibliography}
\end{document}